# Decoupling of the Antiferromagnetic and Insulating States in Tb doped $Sr_2IrO_4$

J. C. Wang[1,2,3], S. Aswartham[1], Feng Ye[2,1], J. Terzic[1], H. Zheng[1], Daniel Haskel[4], Shalinee Chikara[5], Yong Choi[4], P. Schlottmann[6], Radu Custelcean[7], S. J. Yuan[1] and G. Cao[1*]

[1]Center for Advanced Materials and Department of Physics and Astronomy

University of Kentucky, Lexington, KY 40506, USA

[2]Quantum Condensed Matter Division, Oak Ridge National Laboratory,

Oak Ridge, Tennessee 37831, USA

[3]Department of Physics, Renmin University of China, Beijing, China

[4]Advanced Photon Source, Argonne National Laboratory, Argonne IL 60439, USA

[5]National High Magnetic Field Laboratory, Los Alamos National Laboratory,

Los Alamos, NM 87545, USA

[6]Department of Physics, Florida State University, Tallahassee, FL 32306, USA

[7]Chemical Science Division, Oak Ridge National Laboratory,

Oak Ridge, Tennessee 37831, USA

$Sr_2IrO_4$ is a spin-orbit coupled insulator with an antiferromagnetic (AFM) transition at $T_N$=240 K. We report results of a comprehensive study of single-crystal $Sr_2Ir_{1-x}Tb_xO_4$ ($0 \leq x \leq 0.03$). This study found that mere 3% ($x$=0.03) tetravalent $Tb^{4+}$($4f^7$) substituting for $Ir^{4+}$ (rather than $Sr^{2+}$) completely suppresses the long-range collinear AFM transition but retains the insulating state, leading to a phase diagram featuring a decoupling of magnetic interactions and charge gap. The insulating state at x=0.03 is characterized by an unusually large specific heat at low temperatures and an incommensurate magnetic state having magnetic peaks at (0.95, 0, 0) and (0, 0.95, 0) in the neutron diffraction, suggesting a spiral or spin density wave order. It is apparent that Tb doping effectively changes the relative strength of the SOI and the tetragonal CEF and enhances the Hund's rule coupling that competes with the SOI, and destabilizes the AFM state. However, the disappearance of the AFM accompanies no metallic state chiefly because an energy level mismatch for the Ir and Tb sites weakens charge carrier hopping and renders a persistent insulating state. This work highlights an unconventional correlation between the AFM and insulating states in which the magnetic transition plays no critical role in the formation of the charge gap in the iridate.

## I. Introduction

$Sr_2IrO_4$ is a pseudospin-1/2 Heisenberg antiferromagnet (AFM) with a Néel temperature $T_N$ = 240 K [1-4] and AFM exchange coupling approximately 0.1 eV [5]. This system is a prototype for physics driven primarily by a strong interplay of on-site Coulomb U and spin-orbit interactions (SOI) [2-6]. The relativistic SOI, proportional to $Z^2$ (Z is the atomic number), is approximately 0.4 eV in the iridate (compared to ~ 20 meV in 3d materials), and splits the $t_{2g}$ bands into bands with $J_{eff}$ = 1/2 and $J_{eff}$ = 3/2, the latter having lower energy [2-3]. Since the $Ir^{4+}$ ($5d^5$) ions provide five 5d-electrons, four of them fill the lower $J_{eff}$ = 3/2 bands, and one electron partially fills the $J_{eff}$ = 1/2 band where the Fermi level $E_F$ resides. The $J_{eff}$ = 1/2 band is so narrow that even a reduced U (~ 0.50 eV) due to the extended nature of 5d-electron orbitals is sufficient to open a gap (≤ 0.62 eV) supporting a novel insulating state [2-8]. It has become increasingly clear that the new, delicate balance between the relevant energies renders a ground state extremely susceptible to even slight changes of chemical composition [9-12].

$Sr_2IrO_4$ is perhaps the most intensively studied iridate thus far. The distinct energy hierarchy featuring a strong SOI along with its structural and electronic similarities to those of the celebrated $La_2CuO_4$ ($K_2NiF_4$ type, one hole per Ir or Cu ion, pseudospin- or spin-1/2 AFM, etc.) has stimulated a surge of interest in the iridates in recent years. A growing list of theoretical proposals predicts, among many SOI-coupled phenomena, novel topological states and superconductivity via electron or hole doping [5,6,13-15]. A recent study of angle-resolved photoemission exhibits a temperature and doping dependence of Fermi arcs at low temperatures via in situ K doping in cleaved crystal surface of $Sr_2IrO_4$, a phenomenology similar to that of the high-temperature superconducting cuprates [16]. However, superconductivity characterized by zero-resistivity and diamagnetism remains elusive despite extensive experimental efforts.

Nevertheless, a growing body of experimental evidence has shown that even slight electron or hole doping at either Sr or Ir sites leads to a metallic state despite the sizable energy gap (~ 0.62 eV) inherent in $Sr_2IrO_4$. Oftentimes, the AFM state vanishes upon the presence of the metallic state; however, there are exceptions in which the AFM state survives chemical doping at Sr sites and coexists with a doping-induced metallic state **[9, 10, 17, 18]**. Furthermore, recent high-pressure studies of $Sr_2IrO_4$ suggests that the magnetic transition vanishes near 20 GPa **[19]** but the insulating state persists at high pressure up to 55 GPa **[19, 20]**, highlighting an unconventional correlation between the AFM state and insulating gap. Indeed, a signature behavior of $Sr_2IrO_4$ is that transport properties exhibit no discernable anomaly corresponding to the AFM transition at $T_N$=240 K **[1, 18, 21]**, sharply contrasting that of other correlated materials and iridates, such as bilayered $Sr_3Ir_2O_7$ **[22]** and hexagonal $BaIrO_3$ **[23]**. It is not surprising that the unusual character of this SOI-coupled insulator has recently revitalized discussions of Mott, Mott-Hubbard and Slater insulators, particularly, the dependence of charge gap formation on magnetic interactions in $Sr_2IrO_4$ **[24, 25]**. Clearly, a better understanding of the $J_{eff}$=1/2 insulating state and its correlation with the AFM state in $Sr_2IrO_4$ needs to be established.

In this paper, we report results of a comprehensive study of slightly Tb doped $Sr_2IrO_4$ or single-crystal $Sr_2Ir_{1-x}Tb_xO_4$ ($0 \leq x \leq 0.03$). This study utilizes various tools including x-ray absorption near edge structure (XANES) and X-ray absorption fine structure (XAFS), neutron diffraction, and other probes to characterize structural, transport, thermal and magnetic properties of these single crystals. The central finding of this study is that mere 3% ($x$=0.03) tetravalent $Tb^{4+}(4f^7)$ substituting for $Ir^{4+}$ (rather than $Sr^{2+}$) completely suppresses the long-range collinear AFM state but retains an insulating state, leading to a phase diagram featuring a decoupling of magnetic interactions and charge gap. The insulating state at $x$=0.03 exhibits an unusually large specific heat at low temperatures and accompanies an incommensurate magnetic state that is characterized by magnetic peaks at (0.95, 0, 0) and (0, 0.95, 0) in the neutron diffraction, suggesting a spiral or spin density wave order. Slight Tb doping effectively changes the relative strength of the SOI and the tetragonal CEF and enhances the Hund's rule coupling that competes with the SOI, and destabilizes the AFM state, however, there is no concurrent metallic state. This "disentanglement" of charge and magnetic aspects of doped Mott insulators sharply contrasts with the conventional argument where a simultaneous suppression of the magnetic order and charge gap would be anticipated as both are primarily driven by the same force, the Coulomb interaction **[26]**.

## II. Experimental details

The single crystals studied were grown from off-stoichiometric quantities of $SrCl_2$, $SrCO_3$, $IrO_2$ and $Tb_4O_7$ using self-flux techniques **[1, 9-11]**. The size of the single crystals is as large as 2.0 x 2.0 x 0.2 $mm^3$. The structures of $Sr_2Ir_{1-x}Tb_xO_4$ were determined using a Nonius Kappa CCD X-ray diffractometer at the University of Kentucky and a Rigaku X-ray diffractometer equipped with a PILATUS 200K hybrid pixel array detector at Oak Ridge National Laboratory. Full data sets were collected between 100K and 300K, and the structures were refined using the SHELX-97 programs **[27]** and FullProf software **[28]**. Chemical compositions of the single crystals were estimated using energy dispersive X-ray analysis (EDX) (Hitachi/Oxford 3000). The error analysis indicates that the error for Sr and Ir atomic percentage is 0.1% and 3%, respectively, whereas this value for the Tb concentration is ~11%. An example of the EDX results with standard deviation is illustrated in Supplemental Material. Magnetization, specific heat and electrical resistivity were measured using either a Quantum Design MPMS-7 SQUID Magnetometer and/or Physical Property Measurement System with 14-T field capability.

X-ray absorption near edge structure (XANES) and X-ray absorption fine structure (XAFS) measurements were carried out at beamline 4-ID-D of the Advanced Photon Source, Argonne National Laboratory at room temperature. XANES data at the Tb $L_3$ absorption edge were used to determine Tb valence state by comparing leading edge position to reference samples with known valence state: $Tb_4O_7$ with $Tb^{3.5+}$ (i.e., mixed valence with equal amounts of 3+ and 4+ states) and $BaTbO_3$ with $Tb^{4+}$. XANES and XAFS data were collected in fluorescence geometry due to the low Tb content. A 4-element, energy resolving silicon drift diode detector was used to measure the intensity of the Tb $L\alpha_a$ emission as the x-ray energy was scanned through the Tb $L\alpha$ absorption edge. Data were corrected for detector dead time. The reference compounds were in powder form while the Tb doped samples were single crystals oriented in such a way that the electric field of the linearly polarized x-rays was in the $IrO_2$ plane of the tetragonal $Sr_2IrO_4$ structure. XAFS data were collected to 13 Å$^{-1}$ using the same fluorescence geometry. XAFS fits were done using theoretical standards computed with the FEFF6.0 code **[29]**. Prior to fittings, simulations of XAFS data were done by placing Tb atoms at either Ir or Sr sites of the $Sr_2IrO_4$ lattice to determine that Tb occupies Ir sites.

The neutron diffraction was carried out at Oak Ridge National Laboratory using Elastic Diffuse Scattering Spectrometer (CORELLI) at Spallation Neutron Source as well as triple-axis spectrometers HB1 and HB1A at High Flux Isotope Reactor. CORELLI uses a semi-white beam with incident neutron energy ranging from 10 to 200 meV whereas the triple-axis spectrometers HB1 and HB1A utilize an incident energy 13.50 and 14.64 meV, respectively. The temperature control was achieved via a closed cycle refrigerator.

## III. Results and Discussion

Rare earth ions are nominally trivalent but there are a number of exceptions, and Tb is one of them. It can be trivalent $Tb^{3+}$, tetravalent $Tb^{4+}$ or mixed-valent. The trivalent $Tb^{3+}(4f^8)$ ion has a tendency to lose its $8^{th}$ 4f electron to become tetravalent $Tb^{4+}(4f^7)$. We therefore conducted XANES experiments to determine the valence state of Tb. The energy difference for a $2p_{3/2}$ core level electron excitation into empty 5d states between $Tb^{3+}$ and $Tb^{4+}$ ions is about 8 eV as seen in Fig. 1 [29]. A comparison of our XANES results for $x$=0.02 and 0.03 samples with those reported for $Tb^{3+}$ and $Tb^{4+}$ ions [29, 30] clearly indicate the presence of majority tetravalent $Tb^{4+}$ ion in $Sr_2Ir_{1-x}Tb_xO_4$ (see **Fig.1a**). Furthermore, an analysis of our XAFS data concludes that tetravalent $Tb^{4+}$ ions substitute for $Ir^{4+}$ ions rather than $Sr^{2+}$ ions (see more details below). The XAFS data were fitted using FEFF6 theoretical standards [31] generated with the crystal structure of $Sr_2IrO_4$ at room temperature [32-34] and placing a Tb dopant at the Ir site (see more details below). The calculations were done with electric field polarization in the $IrO_2$ plane to match experimental conditions. Data in the k-range of 2-12 Å$^{-1}$ were Fourier transformed into real space and fitted in the range of 1.4 - 4.5 Å. The amplitude reduction factor, $S_0^2$, was found to be 0.81 and an overall $e_0$ shift of 2.41 eV was needed to match the FEFF theory and experiment. The results are illustrated in **Figs.1b** and **1c** where the magnitude and real parts of the complex Fourier transformed XAFS data (black line) and fits (red line) are plotted, respectively. The local Tb-O, Tb-Sr and Tb-Ir distances were found to be expanded by 0.10(1), 0.12(4), 0.05(3) Å relative to the Ir-O, Ir-Sr and Ir-Ir distances in the undoped structure of $Sr_2IrO_4$. This is consistent with results from the XANES measurements where Tb is found to be in the tetravalent state (the ionic radius of $Tb^{4+}$ is 0.13 Å larger than that of $Ir^{4+}$ ions in octahedral environment).

The expansion of the bonding distances of Tb can be attributed to the 4f-electrons: they are localized and have small binding energies, and the S-state of the $Tb^{4+}$ f-shell is spherical and does not favor directional bonds.

We have based our conclusion that Tb dopants occupy Ir sites on the results of simulations of Tb XAFS data using FEFF6.0 theoretical standards (**Fig. 2).** The simulations are remarkably different for Tb doping at Ir and Sr sites. It suffices to inspect the plots in **Fig. 2** to conclude that Tb is located at Ir sites since the simulations in that case reasonably agree with the data even without carrying out any fittings. In contrast, the placement of Tb at Sr sites is inconsistent with the data. The local atomic environments around Ir and Sr sites are very different hence XAFS can easily detect Tb site substitution. For example, for Tb at Ir sites the peaks in **Figs. 2a** are due to oxygen neighbors, Sr neighbors and then Ir neighbors with increasing distance from the Tb site. In comparison, for Tb at Sr sites the first peak is due to oxygen neighbors, Ir neighbors and then Sr neighbors. As seen in these simulations in **Fig. 2**, placing Tb dopants at Ir sites yields perfect fits to the data by accommodating small distortions related to mismatch in ionic radii between Tb and Ir ions. The XAFS data cannot be fitted with a model in which Tb atoms replace Sr atoms.

An unique and important structural feature, which is absent in $La_2CuO_4$, is that $Sr_2IrO_4$ crystallizes in a reduced tetragonal structure (space-group $I4_1/a$) due to a rotation of the $IrO_6$-octahedra about the **c**-axis by ∼11°, resulting in a larger unit cell by √2 x √2 x 2 **[33-37]**. This rotation corresponds to a distorted in-plane Ir-O-Ir bond angle $\theta$ critical to the electronic structure **[9-12].** Slightly substituting $Tb^{4+}$ for $Ir^{4+}$ (up to 3%) retains the tetragonal crystal structure but causes significant changes in the lattice parameters and reduces structural distortions (**Fig.3**). The initial decrease in the unit cell volume V is unusual but it is followed by a sudden increase in V at *x*=0.03, which is anticipated by the increased bonding distances of Tb (**Fig.3a**). Similarly, the c/a ratio decreases initially and then rises at *x*=0.03, roughly tracking the changes in V. Hence, the variation of V=(c/a)a$^3$ with x is predominantly driven by (c/a), i.e. c. It is also remarkable that the rotation of IrO$_6$-octehedra inherent in Sr$_2$IrO$_4$ is considerably reduced so that the Ir-O-Ir bond angle θ increases almost linearly with x from 156.47° at *x*=0 to 160.05° at *x*=0.02 before drops to 158.60° at x=0.03 (**Fig.3b**). These changes have important implications for magnetic properties because of the strong magnetoelastic coupling **[21, 35-39]**.

Impurities are expected to disrupt the itinerant order of the Ir spins, which is based on collective modes, and destroy the spin density waves. Localized moments are much less susceptible to impurities. The tetravalent Tb$^{4+}$(4f$^7$) ion similar to Gd$^{3+}$(4f$^7$) ion is an S-state carrying the total spin and angular momentum S=7/2 and L=0, respectively. The 4f-electrons are localized and have no crystalline field splitting. As shown in **Figs. 4a** and **4b,** 3% Tb doping effectively suppresses the long-range AFM transition T$_N$ from 240 K at x=0 to zero. There is a kink in M(T) for *x*=0 at 100 K that is attributed to a possible rearrangement of the magnetic order and is closely associated with magnetoresistivity **[23]**, magnetoelectric effect **[10]**, unusual muon responses **[37]**. For *x*=0.03, a magnetic hysteresis behavior along with a small kink near 10 K is observed in the *a*-axis magnetic susceptibility χ$_a$ measured using zero-field-cooling (ZFC) and field-cooling sequences (FC) (see **Inset** in **Fig. 4c**); the anomaly is absent in the *c*-axis magnetic susceptibility χ$_c$. This behavior suggests an incommensurate magnetic order, and is confirmed by the neutron diffraction discussed below.

Fitting the magnetic data of χ$_c$ in **Fig. 4c** to a Curie-Weiss law for a temperature range of 1.7-320 K for *x*=0.03 yields the Curie-Weiss temperature θ$_{CW}$ = -1.5 K, consistent with the vanishing T$_N$. A systematic decrease in θ$_{CW}$ or the exchange coupling with *x* closely tracks the decreasing $T_N$, as illustrated in **Fig. 5a**. Extrapolation of $\theta_{CW}(x)$ to $x$ slightly larger than 0.03 suggests a change of sign. The $x$-dependence of $\theta_{CW}$ suggests a vanishing AFM state and an emerging new state with ferromagnetic correlations near $x$=0.03. An increase in the effective moment $\mu_{eff}$ with $x$ is a result of increasing Tb doping that enhances $\mu_{eff}$ from 0.5 for $x$=0 to 1.85 $\mu_B$/f.u. for $x$=0.03 (see **Fig.5a**). With S=7/2, the tetravalent $Tb^{4+}(4f^7)$ ion has an effective moment $\mu_{eff}$ of 7.94 $\mu_B$/Tb. The largely enhanced $\mu_{eff}$ (=1.85 $\mu_B$/f.u.) does not scale well 3% of Tb doping, implying a significant interaction between Ir *5d* and Tb *4f* electrons that intensifies $\mu_{eff}$.

Such a strong *5d-4f* interaction also effectively affects the magnetic anisotropy and ordered moment at low temperatures, and this is evidenced in the isothermal magnetization M(T, H) (**Fig.5b**). For $x$=0, the $a$-axis $M_a$ is more than twice as strong as the $c$-axis $M_c$ because the magnetic moments lie within the basal plane [36, 37]. Upon Tb doping, $M_c$ becomes larger than $M_a$ instead. M(H) at low temperatures is considerably enhanced because of Tb doping. For example, the extrapolated $M_c$ to H=0 for $x$=0.03 is approximately 0.25 $\mu_B$/f.u at T=1.8 K, one order of magnitude stronger than ~ 0.02 $\mu_B$/Ir for $x$=0. The field dependence of M(T, H) at 1.8 K is also suggestive of ferromagnetic (FM) like behavior (see **Fig.5b** as well as **Fig.4**). While for $x$=0 M saturates already at $\mu_o H < 2$ T, for $x$=0.03 the magnetization does not saturate at 14 T.

These changes in the magnetic state are corroborated by results of our neutron diffraction study, as shown in **Fig.6**. With increasing x, a signature magnetic peak at (1,0,2) for the AFM state in x=0 [36, 37] becomes weakened at $x$=0.005 (**Fig.6a**) and eventually vanishes at $x$=0.03. Note that the magnetic peak intensity at $x$=0.005 decreases and the magnetic moment is reduced to 90% of that $x$=0. The disappearance of the sharp magnetic peak associated with the canted antiferromagnetic configuration at $x$=0 is accompanied by an emergent incommensurate magnetic order with wavevectors $q_m$=(0.95, 0, 0) and (0, 0.95, 0). The incommensurate magnetic order becomes better defined at x=0.03 when the higher-*T* background is subtracted (inset of **Fig. 6b**). The intensity of the new peaks is much weaker compared to those at *q*=(1,0,2) for *x*=0. The new peaks exhibit a clear temperature-dependence and evolve into a featureless background above 30 K. The occurrence of the pair of the peaks at (0.95,0,0) and (0, 0.95, 0) implies a possible spiral order with moments along the *c*-axis or an incommensurate spin-density wave as neutron diffraction only probes the moment component perpendicular to the momentum transfer. The spiral order along the c-axis agrees with the stronger $M_c$ at *x*=0.03 (**Fig. 5b**) and is therefore the more likely scenario. It is likely that the magnetic moment of Tb ions, which tends to polarize the magnetic moment of surrounding Ir ions along with it, is ferromagnetically aligned along the *c*-axis or forms magnetic polarons. Generally, a *c*-axis alignment is more energetically favorable when the tetragonal crystal field effect (CFE) is enhanced **[14]**, and specifically, the significantly increased *c/a* ratio in *x*=0.03 (**Fig.3b**) inevitably strengthens the tetragonal CEF, thus favors the *c*-axis alignment for the Ir moments. In addition, the Hund's rule coupling is also enhanced on the Tb sites, further increasing the tendency of a FM interaction along the *c*-axis. A strong competition between the in-plane AFM (due to Ir *5d* electrons) and out of plane FM (due to Tb *4f* electrons) interactions thus accounts for the disappearance of the canted AFM state in *x*=0.03.

The *a*-axis and *c*-axis electrical resistivity, $\rho_a(T)$ and $\rho_c(T)$, systematically reduce with x. $\rho_a(T)$ decreases by nearly four orders of magnitude at low temperatures from ~ $10^6$ Ω cm at x=0 to ~ $10^2$ Ω cm at *x*=0.03, as shown in **Figs.7a** and **7b**. The reduction of $\rho_a(T)$ and $\rho_c(T)$ may be a result of the increased Ir-O-Ir bond angle θ, which makes electron hopping more energetically favorable. However, the insulating state remains at *x*=0.03 with both d$\rho_a$/dT and d$\rho_c$/dT < 0. Indeed, the ratio of ρ(2K)/ρ(300K) for $\rho_a(T)$ merely drops by one order of magnitude, from ~ $10^6$ at $x$ = 0 to ~ $10^5$ at $x$ = 0.03; this ratio for $\rho_c(T)$ at x=0.03 is 2500. A close examination of $\rho_a(T)$ and $\rho_c(T)$ reveals that ρ for x > 0 follows a variable-range hopping (VRH) model, $\rho \sim \exp(1/T)^{1/2}$, for a remarkably wide temperature range, particularly for *x*=0.03. To a lesser extent, ρ for x=0 also follows the VRH behavior (see **Inset** in **Fig.7a** for $\rho_a(T)$), suggesting that the structural distortion may play a significant role. Nevertheless, the VRH behavior signals that Anderson localization, rather than thermal activation, dominates the hoping process with long-range Coulomb repulsions between carriers playing an important role in this regime **[40-42]**. Anderson localization, which is due to disorder in general, may be associated with an energy level mismatch for the Ir and Tb sites that ultimately weakens electron hopping and results in the persistent insulating state.

The specific heat C(T) for 5 < T < 20 K approximately fits a common expression, C(T) = γT + β$T^3$, where the first term arises from the electronic contribution to C(T) and the second term the phonon contribution; γ is usually a measure of the density of states of the conduction states near the Fermi surface and effective mass and β is related to the Debye temperature (see **Fig.8a**). It is therefore intriguing to have relatively large γ in an insulator. The origin of a specific heat linear in T could arise from two-level tunneling centers between two atomic positions (possibly for the O-ions) due to light disorder in the material even for *x*=0 **[17, 18, 21, 43, 44]**. The increase of γ or C(T)/T with x despite the persistent insulating state (**Inset** in **Fig.8a**) is due to the $Tb^{4+}$ spins. Below 5 K, the pronounced upturn in C(T)/T for *x* > 0 and its strong dependence on the magnetic field H that is parallel to the c-axis (see **Fig. 8b**) is likely due to the spin degrees of freedom of the $Tb^{4+}$ S=7/2 states. The field-dependence of C(T)/T exhibits a peak that shifts up with increasing temperature when the field is increased (**Inset** in **Fig. 8b**). This behavior is consistent with weakly correlated $Tb^{4+}$ ions (S=7/2) in a magnetic field. The zero-field entropy $S=\int[C(x=0.03)-C(x=0)](dT/T)$ is 0.63 J/mole K. Converted into the entropy per $Tb^{4+}$- spin we obtain 2.53 $k_B$ which is only slightly larger than the entropy of a S=7/2, i.e. ln(8) $k_B$ = 2.08 $k_B$.

## IV. Summary

A phase diagram is constructed based on the results presented above, as shown in **Fig.8c**. In essence, $Tb^{4+}$ substituting for $Ir^{4+}$ alters the relative strength of the SOI and the tetragonal CEF, and enhances the Hund's rule coupling that competes with the SOI. It is the combined effect of these changes that accounts for the complete suppression of the in-plane AFM state and the occurrence of the spiral order with only 3% of Tb doping. However, an energy level mismatch for the Ir and Tb sites and different symmetry to the different energy levels might depress the carrier hopping between an octahedron containing a Tb ion and one with an Ir ion, potentially preventing the simultaneous emergence of a metallic state. Additional tools such as calculations of Density Functional Theory that could provide more insight into carrier hopping across the lattice would be desirable. Nevertheless, this study provides the considerable empirical evidence that suggests an unconventional correlation between the AFM and insulating state in which the magnetic transition plays no critical role in formation of charge gap in the iridate.

**Acknowledgement**

GC is grateful to Dr. Ribhu Kaul and Dr. Daniel Khomskii for useful discussions. This work was supported by NSF through Grant DMR-1265162 and by the Department of Energy (BES) through Grant No. DE-FG02-98ER45707 (PS). Work at Argonne National Laboratory was supported by the US DOE, Office of Science, Office of Basic Energy Sciences under Contract No. DE-AC02-06CH11357. Work at ORNL was sponsored by the Scientific User Facilities

Division, Office of Basic Energy Sciences, the U.S. Department of Energy. J.C.W. is grateful to the support from Chinese Scholarship Council.

* Corresponding author: cao@uky.edu

**Captions**

**Fig.1. (a)** A comparison of our XANES results for x=0.02 and 0.03 with peaks for $Tb^{3+}$ and $Tb^{4+}$ ions reported in previous studies; **(b)** and **(c)** magnitude and real parts of the complex Fourier transformation (FT) of the XAFS data (black line) and fits (red line), The XAFS data were fitted using FEFF6 theoretical standards generated with the known crystal structure of $Sr_2IrO_4$ at room temperature and placing Tb dopants at Ir sites. Dashed lines indicate the upper bound for the fitting range.

**Fig.2**. Determination of Tb site substitution by comparing XAFS data and simulations: Panels show XAFS data (black dots and lines) and simulations (red lines for Tb at Ir sites, green lines for Tb at Sr sites) for **(a)** magnitude of complex Fourier transform in real space, **(b)** real part of complex Fourier transform in real space, **(c)** XAFS in photoelectron momentum space, and **(d)** back Fourier transform in photoelectron momentum space.

Comparison of XAFS data and simulations: For Tb doping at Ir and Sr sites, Top panel shows data and simulations in real (Fourier transformed space) while lower panels shows data and simulations in photoelectron space. See text for additional details.

**Fig.3.** The Tb concentration dependence of **(a)** the lattice parameters a- and c-axis and unit cell volume V (right scale) and **(b)** the c/a ratio and Ir-O-Ir bond angle θ (right scale).

**Fig.4.** The temperature dependence at $\mu_o H$=0.1 T of the magnetization **(a)** $M_a$ and **(b)** $M_c$ for $0 \leq x \leq 0.03$; and **(c)** $\chi_a$ and $\chi_c$, and $\chi_c^{-1}$ (right scale) for x=0.03; **Inset**: Enlarged $\chi_a$ for both ZFC and FC below 15 K.

**Fi.g.5. (a)** The temperature dependence of $\Delta\chi^{-1}$ for $0 \leq x \leq 0.03$; **Inset**: The Tb concentration x dependence of the Curie-Weiss temperature $\theta_{CW}$ and effective moment $\mu_{eff}$. **(b)** The isothermal magnetization $M_a$ (blue) and $M_c$ (red) up to 14 Tesla for $0 \leq x \leq 0.03$.

**Fig.6.** The Neutron diffraction results: **(a)** Temperature dependence of the peak (1,0,2) for x=0 (black) and x=0.005 (cyan). The intensity is normalized into cross section in the unit of millibarns per unit cell. **Inset**: The rocking curve of (1,0,2) peak for x=0.005 at 5 K (blue) and 250 K (red). **(b)** Temperature dependence of peaks (0.95,0,0) and (0,0.95,0) for x=0.03 in the same unit as in (a). **Upper inset**: the HK0-slice from time-of-flight (TOF) neutron data (on CORELLI). The data at 200K is subtracted as background. The peak around (1,0,0)/(0,1,0) is slightly away from integer positions. **Lower Inset**: High resolution triple axis results (on HB1A) of (H,0,0) scan at 5K (blue) and 50K (red). Consistent with TOF result, the peaks are incommensurate at (0.95,0,0) and (0,0.95,0).

**Fig.7.** The temperature dependence of **(a)** the a-axis resistivity $\rho_a$ and **(b)** the c-axis resistivity $\rho_c$ for $0 \leq x \leq 0.03$.

**Fig.8**. **(a)** The specific heat C(T)/T vs. $T^2$ for $0 \leq x \leq 0.03$ and 1.7<T<18 K; **Inset**: C(T)/T or γ at T=1.8 K vs. x; **(b)** C(T)/T vs. T for x=0.03 at representative magnetic fields, $\mu_o H$, up to 14 T; Inset: C(T)/T or γ vs. $\mu_o H$. **(c)** The phase diagram for $Sr_2Ir_{1-x}Tb_xO_4$ generated based on the data presented above. Note that PM-I stands for a paramagnetic insulator and AFM-I antiferromagnetic insulating state.

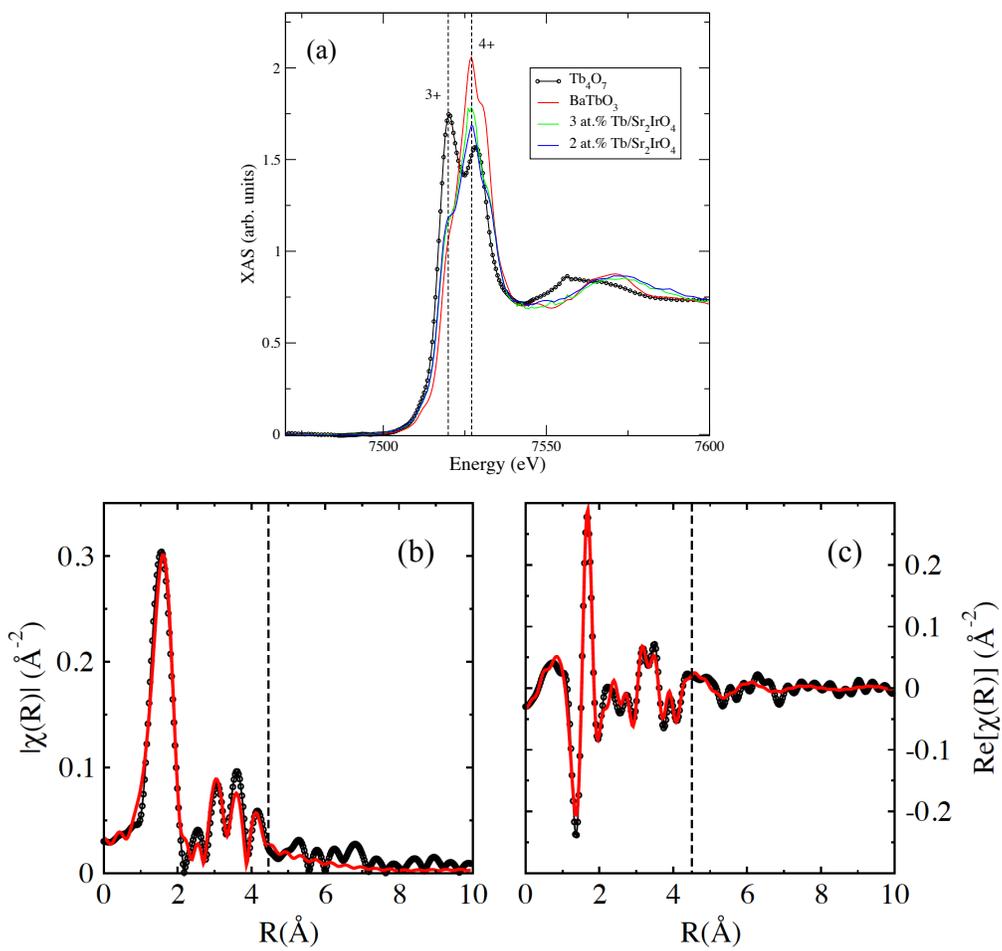

Fig. 1

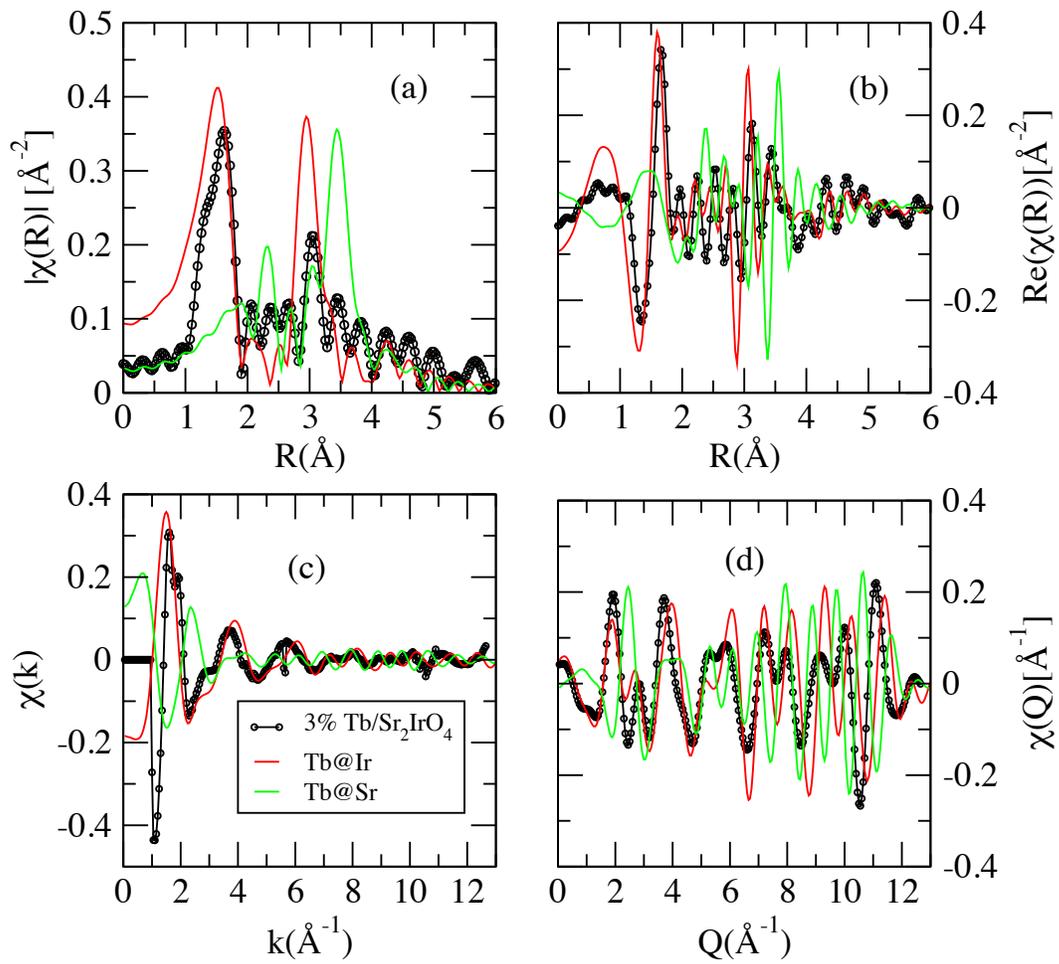

Fig. 2

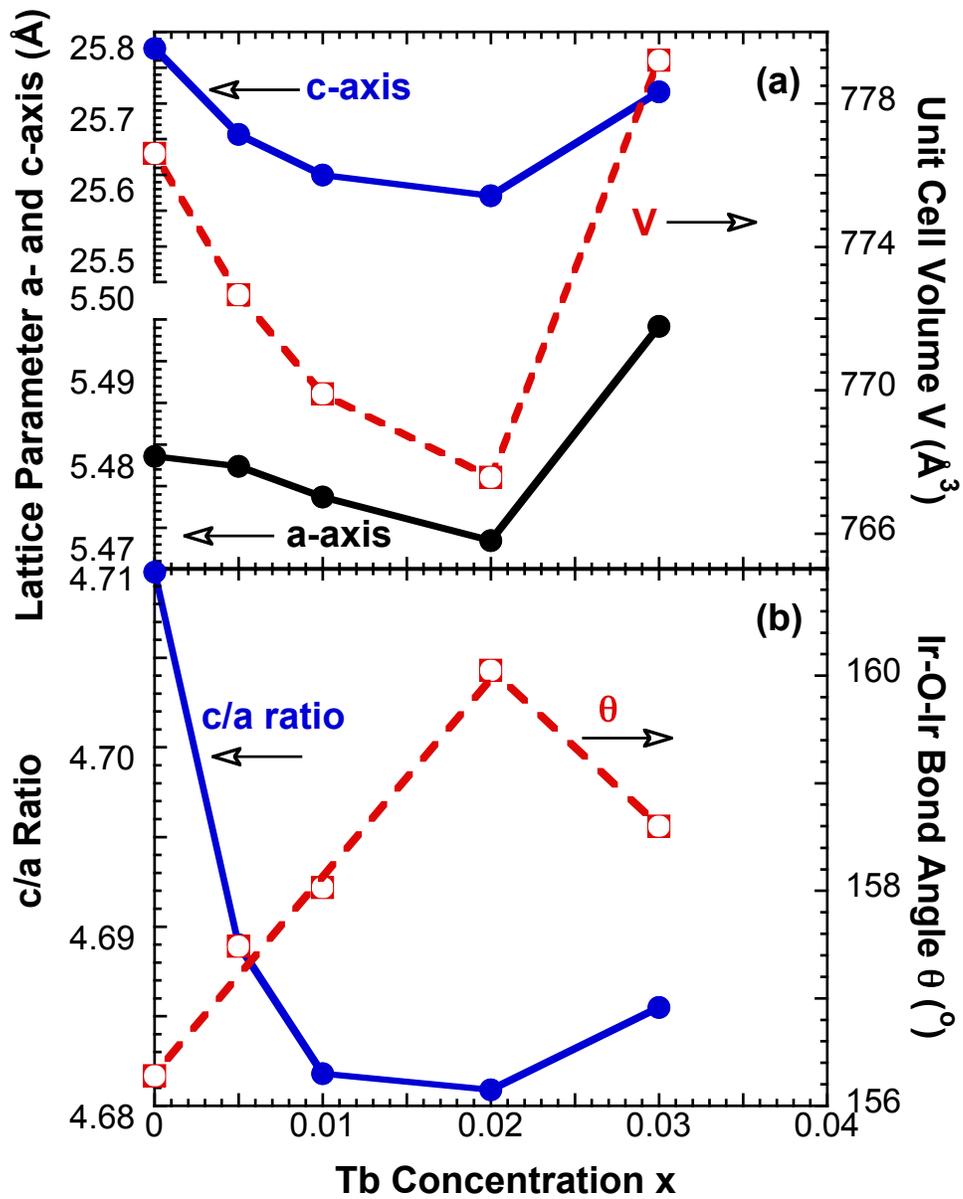

Fig. 3

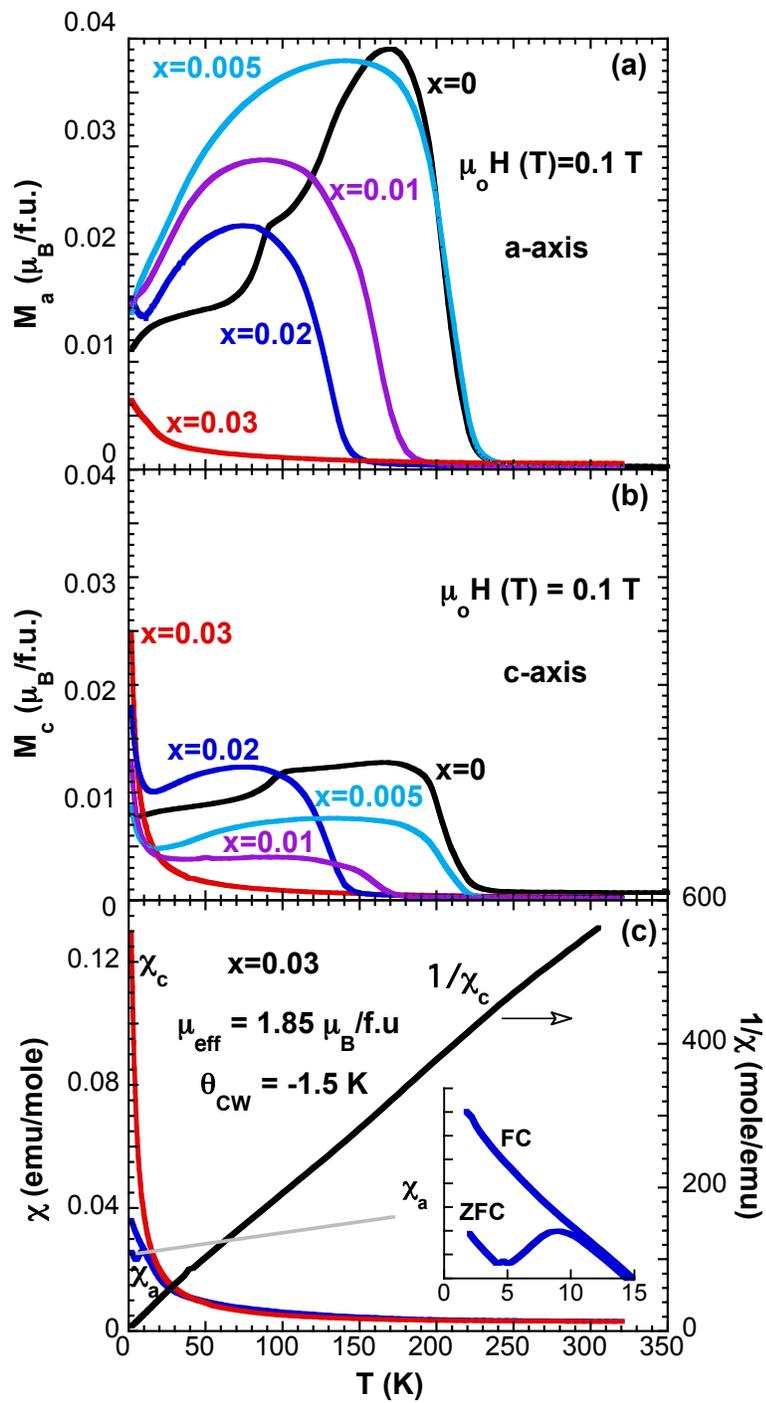

Fig. 4

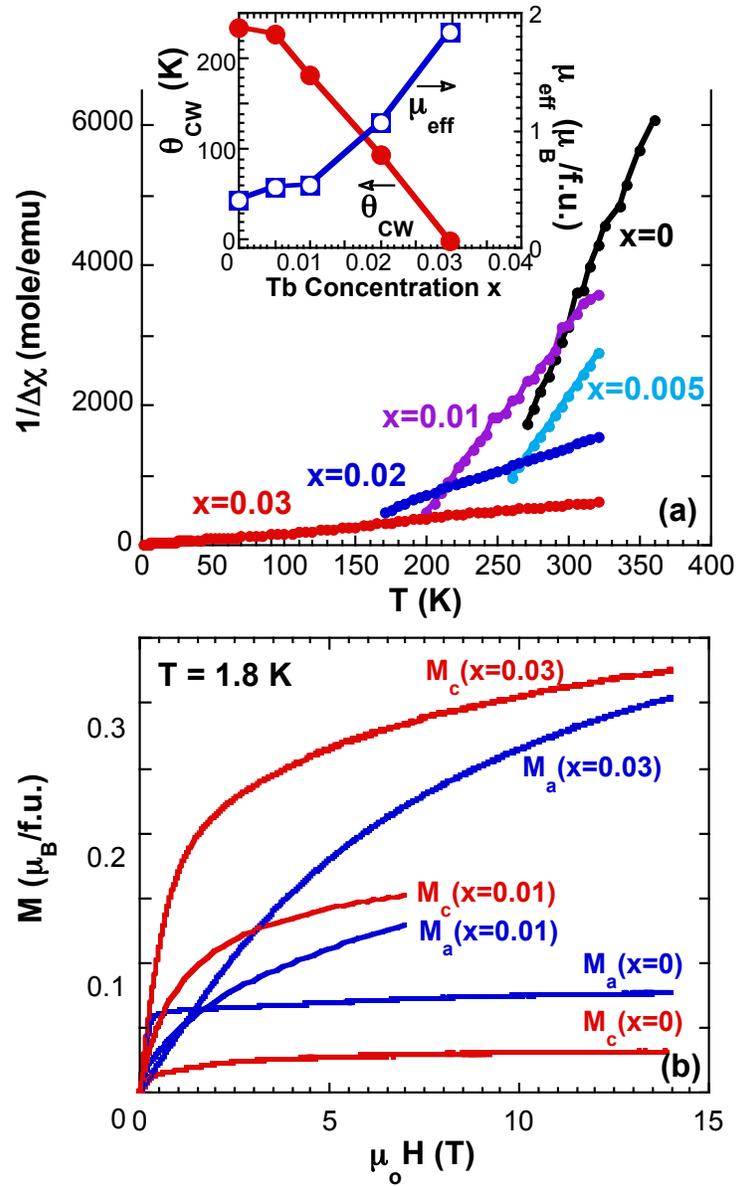

Fig. 5

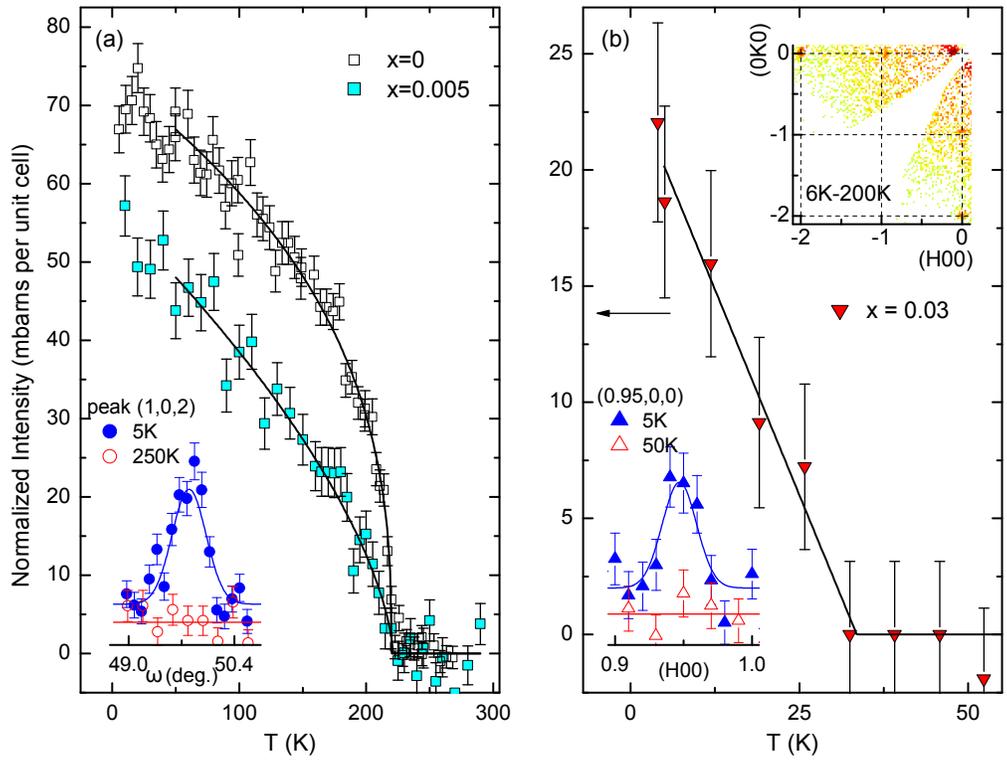

Fig. 6

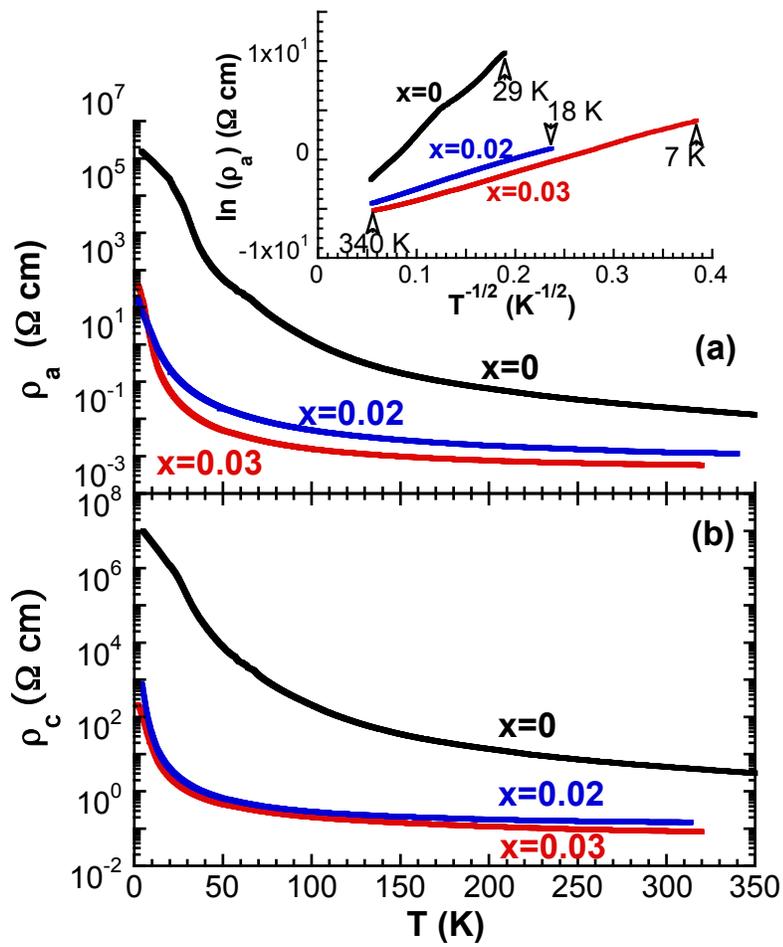

Fig.7

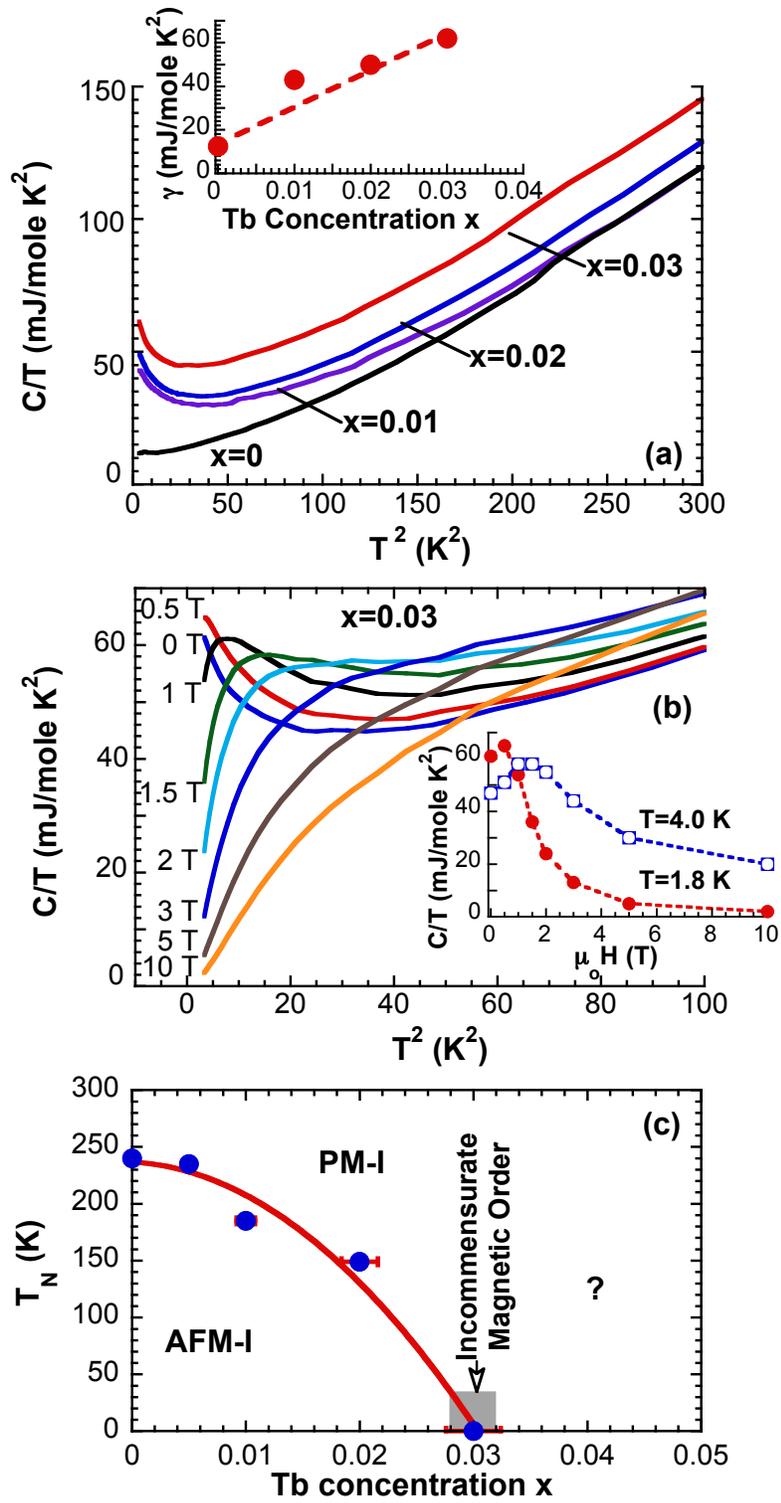

Fig. 8